\documentclass[twocolumn]{aastex63}

\usepackage{amssymb}
\usepackage{amsmath}
\usepackage{natbib}
\usepackage{mathtools}

\shorttitle{Primordial Radius Gap}
\shortauthors{Lee \& Connors}
\graphicspath{{./}{figures/}}

\begin{document}

\title{Primordial Radius Gap and Potentially Broad Core Mass Distributions\\ of Super-Earths and Sub-Neptunes}

\correspondingauthor{Eve J. Lee}
\email{evelee@physics.mcgill.ca}

\author[0000-0002-1228-9820]{Eve J.~Lee}
\affil{Department of Physics, McGill University, Montr\'eal, Qu\'ebec, H3A 2T8, Canada}
\affil{McGill Space Institute, McGill University, Montr\'eal, Qu\'ebec, H3A 2A7, Canada}
\affil{Institute for Research on Exoplanets, Montr\'eal, Qu\'ebec, Canada}

\author{Nicholas J. Connors}
\affil{Department of Physics, McGill University, Montr\'eal, Qu\'ebec, H3A 2T8, Canada}
\affil{School of Computer Science, McGill University, Montr\'eal, Qu\'ebec, H3A 0E9, Canada}

\begin{abstract}

The observed radii distribution of {\it Kepler} exoplanets reveals two distinct populations: those that are more likely to be terrestrials ($\lesssim1.7R_\oplus$) and those that are more likely to be gas-enveloped ($\gtrsim2R_\oplus$). There exists a clear gap in the distribution of radii that separates these two kinds of planets. Mass loss processes like photoevaporation by high energy photons from the host star have been proposed as natural mechanisms to carve out this radius valley. These models favor underlying core mass function of sub-Neptunes that is sharply peaked at $\sim$4--8$M_\oplus$ but the radial-velocity follow-up of these small planets hint at a more bottom-heavy mass function. By taking into account the initial gas accretion in gas-poor (but not gas-empty) nebula, we demonstrate that 1) the observed radius valley is a robust feature that is initially carved out at formation during late-time gas accretion; and 2) that it can be reconciled with core mass functions that are broad extending well into sub-Earth regime. The maximally cooled isothermal limit prohibits cores lighter than $\sim$1--2$M_\oplus$ from accreting enough mass to appear gas-enveloped. The rocky-to-enveloped transition established at formation produces a gap in the radius distribution that shifts to smaller radii farther from the star, similar to that observed. For the best agreement with the data, our late-time gas accretion model
favors dust-free accretion in hotter disks with cores slightly less dense than the Earth ($\sim$0.8$\rho_\oplus$) drawn from a mass function that is as broad as $dN/dM_{\rm core} \propto M_{\rm core}^{-0.7}$.

\end{abstract}

\section{Introduction} 
\label{sec:intro}

In galactic and stellar astronomy, the initial mass function of stars is one of the most fundamental quantity that influences the structural and chemical evolution of the interstellar medium and the galaxy on average. Obtaining an analogous mass function for exoplanets is challenging. Sub-Neptunes and super-Earths dominate the population with many of them at orbital periods beyond $\sim$10 days \citep[e.g.,][]{Fressin13,Petigura13,Burke15}, where we lose sensitivity to measure their masses with e.g., radial velocity surveys \citep[e.g.,][]{Weiss14}. Mass measurements using transit timing variations are available for only a handful of planets in multi-planetary systems, being favorable to those near mean motion resonances \citep[e.g.,][]{Wu-ttv13,Hadden14}.

Theoretically, \citet{Malhotra15} derived a log-normal distribution of total mass (i.e., core + envelope mass) function peaked at $\sim$4--10$M_\oplus$ using the observed period ratio distribution and applying the condition for dynamical stability given by Hill spacing. \citet{Wu19} searched for a log-normal distribution of core masses that best-fits photoevaporation model to the observed distribution of planetary radii. They argued that a mass distribution sharply peaked at $\sim$8$M_\oplus(M_\star/M_\odot)$ was necessary to reproduce the shape of the ``radius valley'', a gap in the radius distribution at $\sim$1.3--1.6 $R_\oplus$ predicted by mass loss theory \citep{Owen13} and later confirmed by the California-Kepler Survey \citep{Fulton17,Fulton18} and asteroseismology \citep{vanEylen18}. \citet{Rogers20} performed a more sophisticated hierarchical inference analysis fitting photoevaporation model to the observed radius-period distribution and concluded a similarly peaked mass distribution (with the median at $\sim$4$M_\oplus$) is required. 

Such high masses are at odds with the radial velocity follow-up of {\it Kepler} planets which reports peak masses as low as $\sim$1$M_\oplus$ \citep{Weiss14}. Furthermore, the true radius/mass distribution may be more bottom-heavy than previously thought \citep{Hsu19}.

In this paper, we assess whether a power-law core mass distribution that extends to the sub-Earth masses is consistent with the observed radius distribution as well as the shape of the gap in the radius-period space. Instead of assuming a distribution of initial envelope mass fraction that is independent of core mass, we calculate the expected envelope mass from nebular accretion in the late stages of disk evolution: a gas-poor environment deemed favorable for preventing runaway gas accretion to ensure the formation of super-Earths and sub-Neptunes \citep{Lee14,Lee16}. 

Section \ref{sec:methods} outlines the basic physical ingredients for gas accretion and photoevaporative mass loss, and the model results are presented in Section \ref{sec:results}. We summarize, discuss the implications, and conclude in Section \ref{sec:disc_concl}.

\section{Methods}
\label{sec:methods}

\subsection{Underlying core mass distribution}

\begin{figure}
    \centering
    \includegraphics[width=0.45\textwidth]{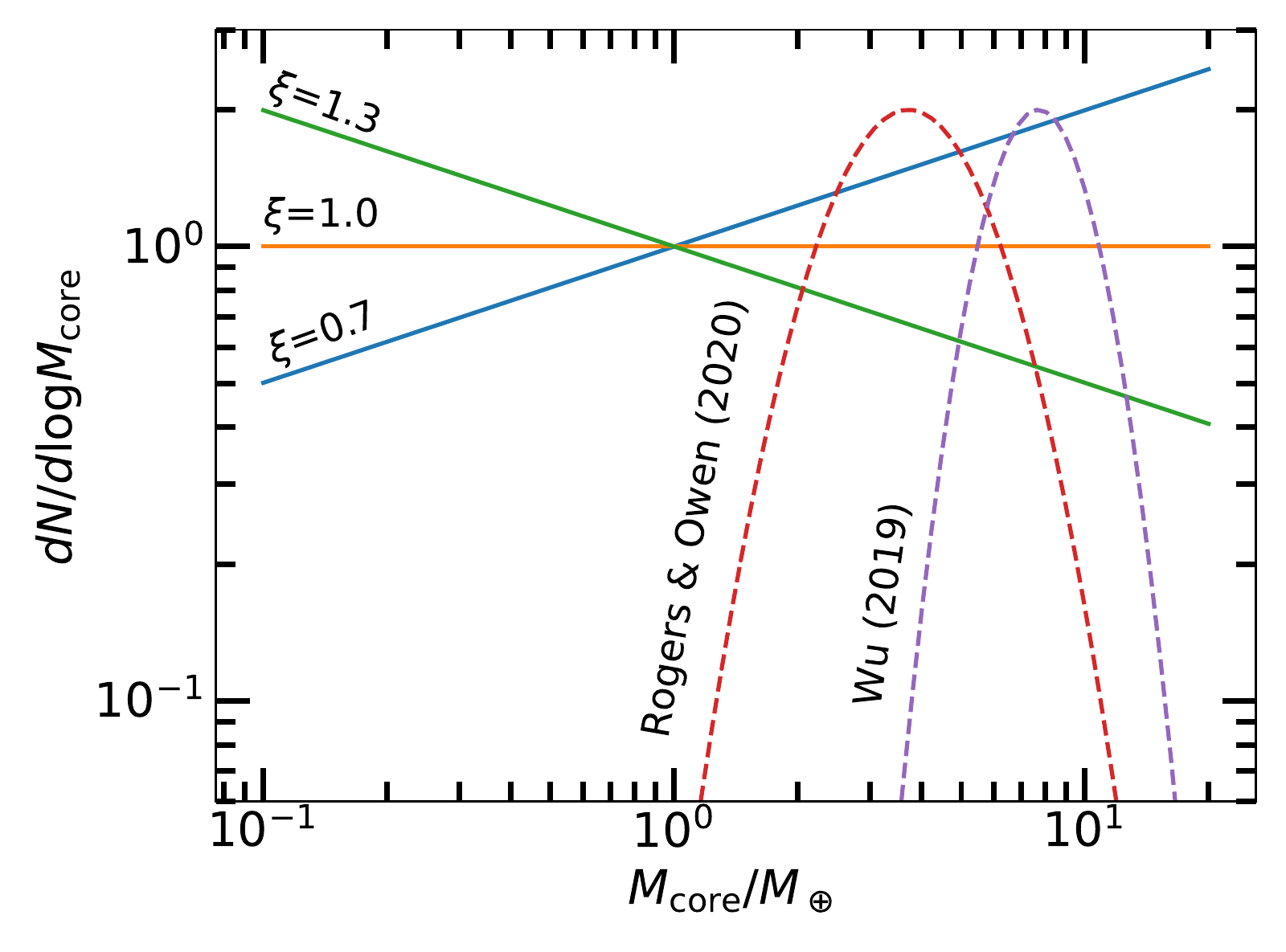}
    \caption{Comparison of power-law core mass distributions used in this paper and best-fit lognormal distributions from \citet{Wu19} and \citet{Rogers20}, their Model I. The non-parametric core mass functions favored by \citet{Rogers20} are broader but still feature a sharp fall-off at small core masses, unlike smooth power-law distributions explored in this paper. Normalizations are arbitrary and are adjusted for better visualization.}
    \label{fig:dNdlogM_comp}
\end{figure}

We begin with the ansatz that the underlying sub-Neptune/super-Earth core mass distribution follows a power-law distribution:
\begin{equation}
    \label{eq:dNdM_pow}
    \frac{dN}{d\log M_{\rm core}} \propto M_{\rm core}^{1-\xi},
\end{equation}
where $M_{\rm core}$ is the mass of the core and we choose $\xi \in [0.7, 1.0, 1.3]$; $\xi=0.7$ is the best-fit power-law slope to the distribution of peak posterior masses of sub-Neptunes from the radial-velocity follow-up by \citet{Marcy14}. Given that radial velocity measurements are biased against low mass objects, we do not choose $\xi$ lower than 0.7. We note that $\xi=0.7$ is top-heavy, $\xi=1.0$ is neutral, and $\xi=1.3$ is bottom-heavy. We also experimented with exponential distribution in linear and logarithm of $M_{\rm core}$ and found them to provide poor match to the data. The minimum and the maximum core masses are set to 0.01$M_\oplus$ and 20$M_\oplus$, respectively. Figure \ref{fig:dNdlogM_comp} demonstrates the difference between the assumed power-law mass distributions with respect to the best-fit lognormal distributions presented in literature.

\subsection{Initial envelope mass fraction}
\label{ssec:init}

For each core, its initial envelope mass fraction is calculated using the analytic scaling relationship derived by \citet{Lee15} appropriate for gas accretion by cooling (equivalent to Phase II of the core accretion theory, \citealt{Pollack96}; see also \citealt{Ginzburg16}). We modify the expressions for the weak dependence on the nebular density \citep{Lee16} and for the expected decrease in the bound radius due to three-dimensional hydrodynamic effects \citep{Lambrechts17,Fung19}. Shrinking the outer bound radius decreases the rate of accretion in a linear fashion (\citealt{Lee14}; see also \citealt{Alidib20} for understanding this effect in terms of entropy delivery). We verified that the expressions we provide here match the numerical calculations. 

First, cores need to be sufficiently massive to accrete gas. Their gravitational sphere of influence must encompass the cores themselves so that the speed of gas is below the cores' surface escape speed. We calculate the envelope mass only for cores that satisfy
\begin{align}
    R_{\rm core} &\leq R_{\rm out} \equiv f_R\,{\rm min}(R_{\rm Hill}, R_{\rm Bondi}) \nonumber \\
    M_{\rm core} &\geq 0.02\,M_\oplus\,\left(\frac{T_{\rm disk}}{1000\,K}\right)^{4/3},
\end{align}
where $R_{\rm core} \equiv R_\oplus (M_{\rm core}/M_\oplus)^{1/4}$ \citep{Valencia06}, $R_{\rm out}$ is the outer radius of the bound envelope, $f_R < 1$ is a numerical factor that takes into account the effect of three-dimensional advective flows, $R_{\rm Hill}$ is the Hill radius, $R_{\rm Bondi}$ is the Bondi radius, $T_{\rm disk} = 1000\,{\rm K}\,f_T(a/0.1\,{\rm AU})^{-3/7}$ is the disk temperature, $a$ is the orbital distance, and $f_T$ is a numerical coefficient and a free parameter. We note that for these small cores, $R_{\rm Bondi} < R_{\rm Hill}$ inside 1 AU. We have implicitly assumed a passively heated disk (i.e., heating dominated by stellar irradiation). In Sections \ref{ssec:prim_Rvalley} and \ref{ssec:Tdisk}, we will explore the effect of adopting temperature profiles relevant for active disks (i.e., heating dominated by viscous accretion).

For dusty accretion, the envelope mass fraction
\begin{align}
    \frac{M_{\rm env}}{M_{\rm core}} &= 0.06\,f_R\left(\frac{M_{\rm core}}{5\,M_\oplus}\right)^{1.8}\left(\frac{t}{1\,{\rm Myrs}}\right)^{0.4} \nonumber \\
    & \times \left(\frac{\Sigma_{\rm gas}}{2000\,{\rm g}\,{\rm cm}^{-2}}\right)^{0.12}\left(\frac{0.02}{Z}\right)^{0.4}\left(\frac{\mu}{2.37}\right)^{3.4}
    \label{eq:dusty_accr}
\end{align}
where $M_{\rm env}$ is the mass of the gaseous envelope, $t$ is the accretion time, $\Sigma_{\rm gas} = 1.3\times 10^5\,{\rm g}\,{\rm cm}^{-2}\,f_{\rm dep}\,(a/0.2\,{\rm AU})^{-1.6}$ is the local disk gas surface density \citep{Chiang13}, $f_{\rm dep}$ is the disk gas depletion factor, $Z$ is the envelope metallicity, and $\mu$ is the envelope mean molecular weight. Similarly, for dust-free accretion,
\begin{align}
    \frac{M_{\rm env}}{M_{\rm core}} &= 0.25\,f_R\left(\frac{M_{\rm core}}{5\,M_\oplus}\right)^{1.1}\left(\frac{t}{1\,{\rm kyrs}}\right)^{0.4} \left(\frac{200\,{\rm K}}{T_{\rm disk}}\right)^{1.5} \nonumber \\
    & \times \left(\frac{\Sigma_{\rm gas}}{4\times 10^5\,{\rm g}\,{\rm cm}^{-2}}\right)^{0.12} \left(\frac{0.02}{Z}\right)^{0.4}\left(\frac{\mu}{2.37}\right)^{2.2}.
    \label{eq:df_accr}
\end{align}
We express equation \ref{eq:df_accr} with the disk temperature $T_{\rm disk}$. More precisely, the relevant temperature is that at the envelope radiative-convective boundary. The outer layers of dust-free envelopes are nearly isothermal so adopting $T_{\rm disk}$ obtains the same answer. We note that equations \ref{eq:dusty_accr} and \ref{eq:df_accr} have been adjusted for $R_{\rm core} \propto M_{\rm core}^{1/4}$ compared to \citet{Lee15} who used $R_{\rm core} \propto M_{\rm core}^{1/3}$; this adjustment makes no significant difference in our conclusions.

Throughout this paper, $Z=0.02$ (solar metallicity), $\mu = 2.37$, and $t$ is drawn from a logarithmically uniform distribution that range 0.01 and 1 Myr, consistent with the late-time formation scenario \citep{Lee16}. The log-uniform distribution is chosen to account for core formation by a series of collisional mergers whose doubling timescales lengthen with time \citep[e.g.,][]{Dawson15}. Motivated by Figure 11 of \citet{Fung19}, we explore $f_R = 0.1$ and 0.2. We choose $f_{\rm dep} = 0.001$ throughout, prompted by the required level of gas depletion to reproduce the observed peaks in period ratios just outside of first order mean-motion resonances \citep{Choksi20}.

For a given core mass, the maximum possible envelope mass that can be accreted is given by a fully isothermal profile \citep[e.g.,][]{Lee15}. No cores are allowed to accrete more than this maximally cooled isothermal mass:
\begin{equation}
    \label{eq:Miso}
    M_{\rm iso} = 4\pi\rho_{\rm disk}\int^{R_{\rm out}}_{R_{\rm core}} r^2\,{\rm Exp}\left[\frac{GM_{\rm core}}{c_{s,disk}^2}\left(\frac{1}{r}-\frac{1}{R_{\rm out}}\right)\right]dr,
\end{equation}
where $\rho_{\rm disk} \equiv \Sigma_{\rm gas}\Omega/c_{\rm s,disk}$ is the local nebular volumetric density, $\Omega$ is the Keplerian orbital frequency, $c_{\rm s,disk} = kT_{\rm disk}/\mu m_H$ is the local disk sound speed, $k$ is the Boltzmann constant, and $m_H$ is the mass of the hydrogen atom. The nebular mean molecular weight $\mu$ is assumed to be the same as that of the envelope.

\subsection{Estimating radii}

While the masses of sub-Neptunes are dominated by the cores, their radii are largely determined by their envelope mass fraction \citep{Lopez14}. We follow closely the procedure devised by \citet{Owen17} in converting envelope mass fractions to radii. Only the essential elements are shown here.

First, we assume that after the disk gas is completely dissipated and planets are laid bare to stellar insolation, their outer layers become isothermal and volumetrically thin ($\sim$6 scale heights above the radiative-convective boundary; \citealt{Lopez14}). From the density profile given by the inner adiabat
\begin{equation}
    \rho(r) \simeq \rho_{\rm rcb}\left[\nabla_{\rm ad}\frac{GM_{\rm core}}{c_s^2}\left(\frac{1}{r}-\frac{1}{R_{\rm rcb}}\right)\right],
\end{equation}
the total envelope mass
\begin{equation}
    \label{eq:Menv}
    M_{\rm env} \simeq 4\pi\rho_{\rm rcb}R_{\rm rcb}^3\left(\nabla_{\rm ad}\frac{GM_{\rm core}}{c_s^2 R_{\rm rcb}}\right)^{1/(\gamma-1)}I_2,
\end{equation}
where $\rho_{\rm rcb}$ is the density at the radiative-convective boundary (rcb), $\nabla_{\rm ad} \equiv (\gamma-1)/\gamma$ is the adiabatic gradient, $\gamma$ is the adiabatic index of the interior, $G$ is the gravitational constant, $c_s \equiv kT_{\rm eq}/\mu m_H$ is the sound speed evaluated at the location of the planet, $T_{\rm eq} \equiv T_{\rm eff,\odot}(R_\odot/a)^{0.5}$ is the equilibrium temperature of the planet, $R_{\rm rcb}$ is the radius at the radiative-convective boundary, and $I_2$ is the structure integral that follows the form
\begin{equation}
    \label{eq:strucI}
    I_n \equiv \int^1_{R_{\rm core}/R_{\rm rcb}} x^n (x^{-1}-1)^{1/(\gamma-1)} dx.
\end{equation}

To eliminate $\rho_{\rm rcb}$, we use temperature gradient at the rcb so that
\begin{equation}
    \label{eq:rhorcb0}
    \rho_{\rm rcb} = \frac{64\pi\sigma_{\rm sb}\mu m_H}{3k\kappa}\nabla_{\rm ad}\frac{GM_{\rm core}T_{\rm eq}^3}{L},
\end{equation}
where $\sigma_{\rm sb}$ is the Stefan-Boltzmann constant, $\kappa \equiv 10^C \rho_{\rm rcb}^\alpha (k/\mu m_H)^\alpha T_{\rm eq}^{\alpha+\beta}$ is the opacity at the rcb, and $L$ is the cooling luminosity, which can be written as
\begin{equation}
    \label{eq:L}
    L \simeq \frac{GM_{\rm core}M_{\rm env}}{\tau_{\rm KH}R_{\rm rcb}}\frac{I_1}{I_2},
\end{equation}
where $\tau_{\rm KH}$ is the Kelvin-Helmholtz cooling time of the envelope, and $I_1$ again follows the structure integral given by equation \ref{eq:strucI}. We vary $\tau_{\rm KH} \in$ (100, 300) Myrs. The former choice provides a good agreement between the analytically derived planetary radius presented here with the numerically computed thermally evolving sub-Neptunes by \citet{Lopez14} at 100 Myrs. The latter choice agrees well with the numerical solutions of \citet{Lopez14} at 1 Gyrs.
Substituting equation \ref{eq:L} into equation \ref{eq:rhorcb0},
\begin{align}
    \label{eq:rhorcb1}
    \rho_{\rm rcb}^{1+\alpha} &= \frac{64\pi\sigma_{\rm sb}\mu m_H}{3k} \nabla_{\rm ad} 10^{-C} \left(\frac{\mu m_H}{k}\right)^{\alpha}T_{\rm eq}^{3-\alpha-\beta}\frac{I_2}{I_1} \nonumber \\
    &\times \frac{\tau_{\rm KH}}{M_{\rm env}}\left(\frac{R_{\rm rcb}}{R_{\rm core}}\right)R_{\rm core}.
\end{align}
By re-arranging equation \ref{eq:Menv}, we find another equation for $\rho_{\rm rcb}$:
\begin{align}
    \label{eq:rhorcb2}
    \rho_{\rm rcb} &= \frac{M_{\rm env}}{4\pi}\left(\frac{R_{\rm rcb}}{R_{\rm core}}\right)^{-3+1/(\gamma-1)}R_{\rm core}^{-3+1/(\gamma-1)} \nonumber \\ &\times \left(\nabla_{\rm ad}\frac{GM_{\rm core}}{c_s^2}\right)^{1/(1-\gamma)}I_2^{-1}.
\end{align}
We numerically solve for $R_{\rm rcb}/R_{\rm core}$ that obtains $\rho_{\rm rcb}$ satisfying both equations \ref{eq:rhorcb1} and \ref{eq:rhorcb2}, using the \texttt{root\_scalar} function from \texttt{SciPy optimize} package. 
Throughout the paper, we adopt $\gamma=7/5$,\footnote{We note that at formation, the inner adiabat follows more closely $\gamma = 1.2$ as the energy is spent on dissociating hydrogen molecules. It is expected that $\gamma$ approaches 7/5 as the envelope cools below the dissociation temperature $\sim$2500 K but this is yet to be verified with detailed, self-consistent calculation that tracks planets from their formation through post-disk evolution.}, $C=-7.32$, $\alpha=0.68$, and $\beta=0.45$ \citep{Rogers10}.\footnote{These values are obtained by fitting to the opacities tabulated by \citet{Freedman08}, which is designed for dust-free atmospheres. In the absence of post-disk pollution by nearby small grains or giant impact, it is reasonable to consider the upper envelope to be drained out of grains (the gravitational settling timescale of a micron-sized grain is about 1 Myr).} 
To save computation time, we set $R_{\rm rcb}/R_{\rm core} = 1$ for any $M_{\rm env}/M_{\rm core}$ that gives $R_{\rm rcb}/R_{\rm core} < 1.05$, motivated by the $\sim$5\% error in {\it Kepler} transit depth measurements \citep[e.g.,][]{Fulton18}. This limit can be found easily by taking the limit of $R_{\rm rcb}/R_{\rm core} \longrightarrow 1$ and confirming numerically:
\begin{equation}
    \left.\frac{M_{\rm env}}{M_{\rm core}}\right|_{\rm min} = 4.4\times 10^{-5} \left(\frac{M_{\rm core}}{M_\oplus}\right)^{0.74} \left(\frac{a}{0.42\,{\rm au}}\right)^{0.44},
\end{equation}
for $\tau_{\rm KH}=100$ Myrs. The numerical prefactor changes to $7\times 10^{-5}$ for $\tau_{\rm KH}=300$ Myrs.

The photospheric radius---the observable---is a few scale heights above $R_{\rm rcb}$. Correction for the photosphere is made using
\begin{equation}
    R_{\rm phot} = R_{\rm rcb} + \ln\left(\frac{\rho_{\rm rcb}}{\rho_{\rm ph}}\right)\frac{kT_{\rm eq}}{\mu m_H g}
\end{equation}
where $\rho_{\rm ph} = (2/3) \mu m_H g / k T_{\rm eq} \kappa$ is the density at the photosphere and $g \equiv GM_{\rm core}/R_{\rm rcb}^2$ is the surface gravity.

\begin{figure}
    \centering
    \includegraphics[width=0.45\textwidth]{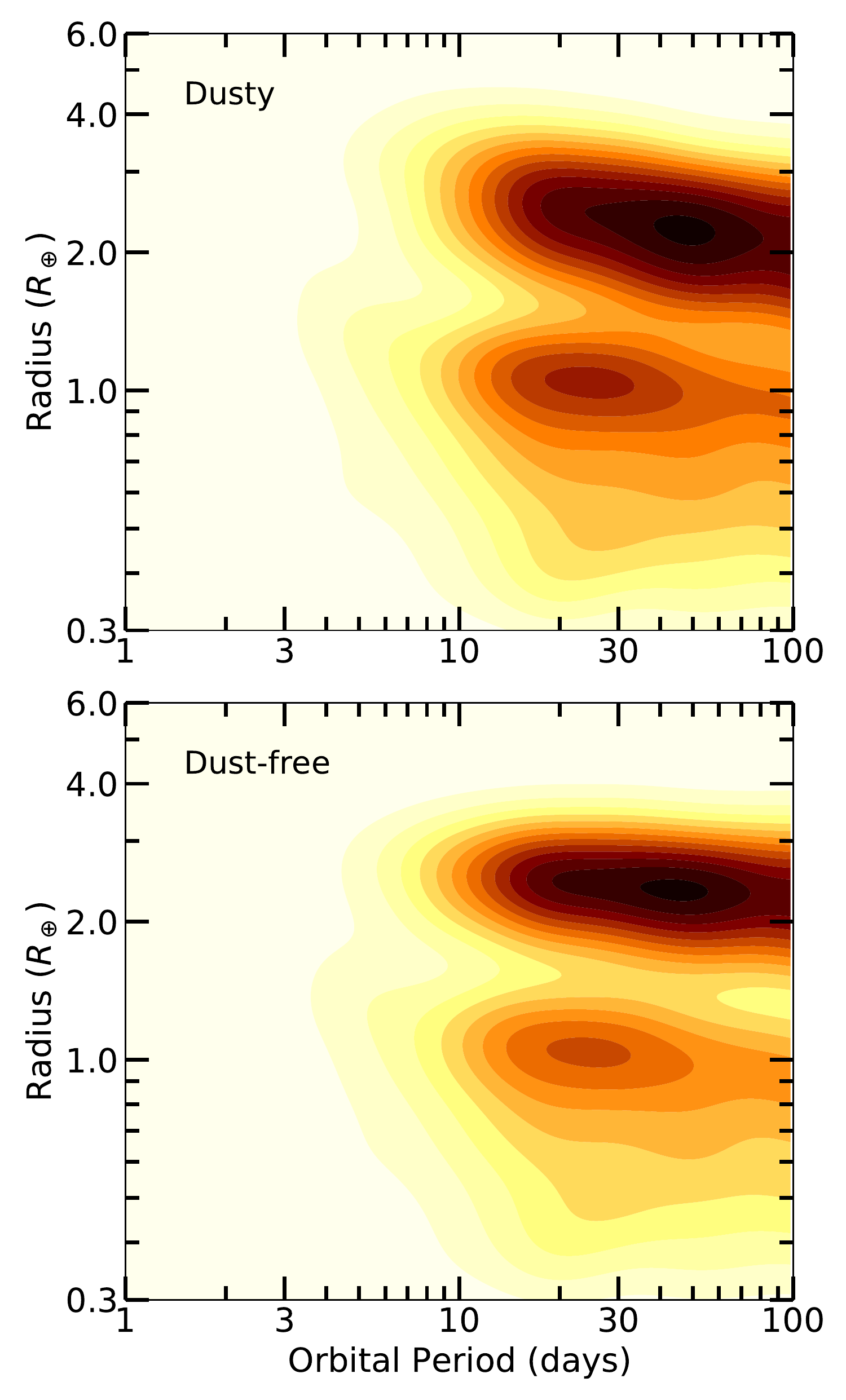}
    \caption{Primordial photometric radius vs. orbital period distribution with $f_R=0.1$, $f_T=2$, $\xi=0.7$, $\rho_c = \rho_\oplus$, and $\tau_{\rm KH} = 100$ Myrs. We smooth the model data using Gaussian kernels with Scott's Rule for bandwidth selection (\texttt{SciPy}'s \texttt{gaussian\_kde} function). Gas accretion is assumed to proceed for 1 Myr in a nebula depleted by three orders of magnitude with respect to the standard solar value ($f_{\rm dep}=0.001$). The distinction between the two population of planets is more apparent in dust-free models. For both dusty and dust-free accretion, the rocky-to-enveloped transition shifts to smaller radii at longer orbital periods.}
    \label{fig:prim_RvP}
\end{figure}

\subsection{Envelope mass loss}

Once the disk gas dissipates and the planets are laid bare to stellar insolation, those that are closest to the star are expected to lose their gaseous envelopes by hydrodynamic winds, either by photoevaporation \citep[e.g.,][]{Owen13} or by internal heat \citep[e.g.,][]{Ikoma12,Owen16,Ginzburg18}. 
The key difference between the two mechanisms 
that bear on observations
is the source of insolation: whereas the former depends on the high-energy flux, the latter depends on the bolometric flux. Using this difference to validate one process over another remains to be carried out.
There is a discernible shift in the position of the gap towards larger radius around more massive host stars \citep{Fulton18,Cloutier20,Berger20} . To reproduce this feature, photoevaporative model requires stellar-mass dependent core mass distribution \citep{Wu19} whereas this is a natural prediction of Parker wind, core-powered envelope mass loss model \citep{Gupta20}.
For solar-type stars, the two mechanisms predict similar location and shape of the gap in the radius-period distribution. Since the goal of this paper is to assess the likelihood of broad sub-Neptune core mass functions for a fixed mass of the host star, we limit our analysis to photoevaporative mass loss for simplicity. We discuss potential effects of varying stellar masses in Section \ref{sec:disc_concl}.

Following \citet{Owen17}, we evolve the envelope mass over 1 Gyrs according to the energy-limited mass loss \citep[e.g.,][]{Lopez13}
\begin{equation}
    \dot{M}_{\rm env} = -\eta \frac{L_{\rm HE} R_{\rm phot}^3}{4a^2G(M_{\rm core} + M_{\rm env})}
\end{equation}
where $\eta = 0.1$ is the mass loss efficiency factor, and $L_{\rm HE}$ is the high-energy luminosity of the star \citep[e.g.,][]{Ribas05,Jackson12}
\begin{equation}
    L_{\rm HE} = \begin{cases}
    10^{-3.5}\,L_\odot & t < 100\,{\rm Myrs},\\
    10^{-3.5}\,L_\odot\left(\frac{t}{100\,{\rm Myrs}}\right)^{-1.5} & t \geq 100\,{\rm Myrs}.
    \end{cases}
\end{equation}

Orbital periods are drawn from the empirical distribution following \citet{Petigura18}
\begin{equation}
    \frac{dN}{d\log P} = 0.52\,P^{-0.1} \left[1-{\rm Exp}\left(-\left(\frac{P}{11.9\,{\rm days}}\right)^{2.4}\right)\right]
\end{equation}
and then converted to orbital distance assuming solar mass host star.

\begin{figure}
    \centering
    \includegraphics[width=0.45\textwidth]{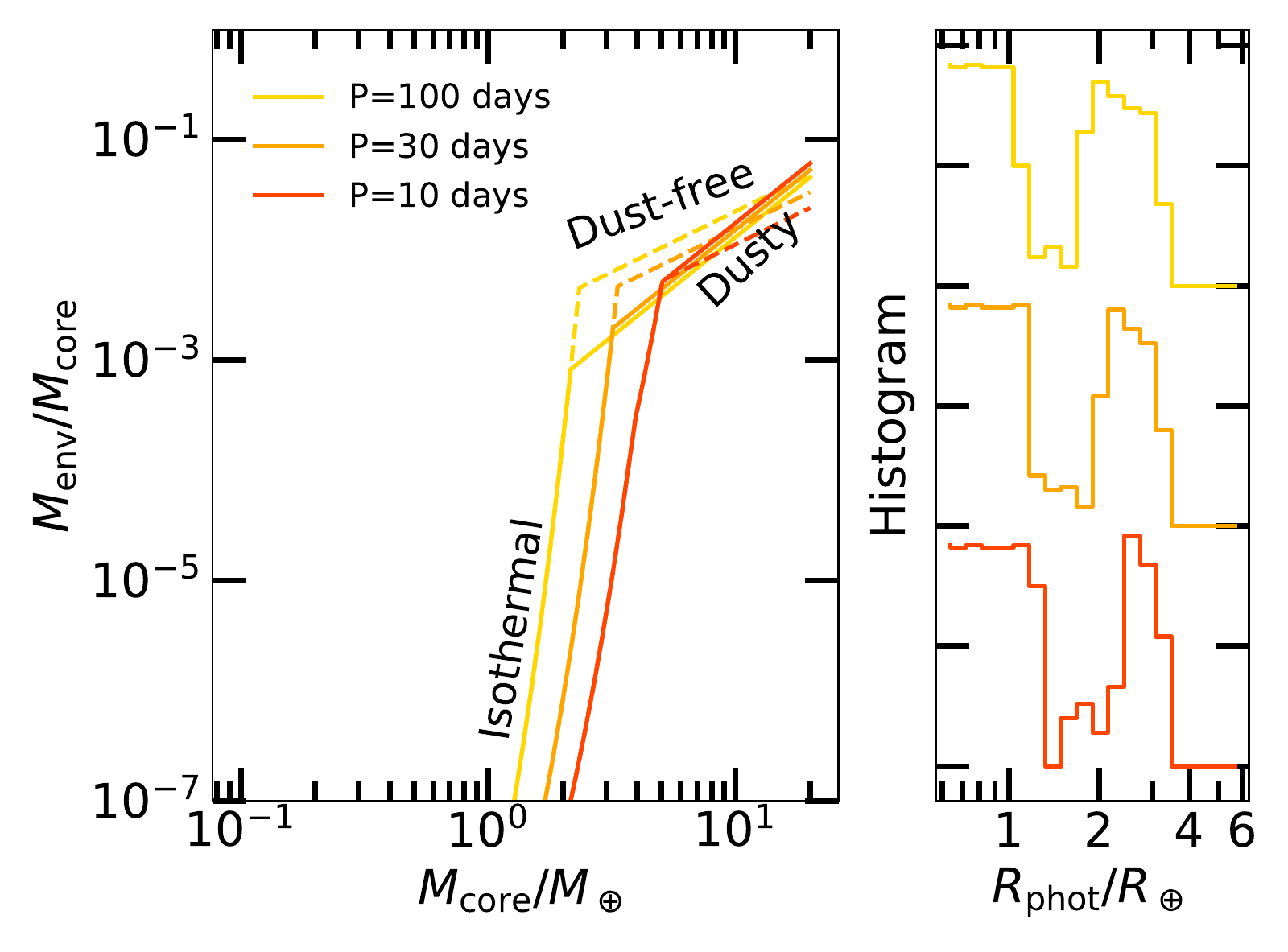}
    \caption{The primordial rocky to enveloped transition as a function of orbital period. Left: envelope mass fraction vs.~core mass after 1 Myr of accretion for $f_R=0.1$, $f_T=2$, $\rho_c = \rho_\oplus$, and $f_{\rm dep}=0.001$. The maximally cooled isothermal limit truncates the gas accretion curves at $\sim$1--4$M_\oplus$. At longer orbital periods, the isothermal mass rises and so the truncation core mass shrinks. Right: histogram of photometric radii for dust-free accretion. The deep gap seen in the histogram coincides with the isothermal truncation mass shown in the left panel.}
    \label{fig:isocool}
\end{figure}

\section{Results}
\label{sec:results}

\subsection{Primordial Radius Valley from Late-time Gas Accretion}
\label{ssec:prim_Rvalley}

We first show that late-time gas accretion alone produces a gap in the radius-period distribution (see Figure \ref{fig:prim_RvP}). The amount of gaseous envelope a core can accrete drops sharply below $\sim$2$M_\oplus$ as their gas masses are limited by the maximally cooled isothermal state. The exponential dependence of this isothermal envelope mass on the core mass (equation \ref{eq:Miso}) implies a bimodal distribution of envelope mass fractions and therefore a bimodal distribution of radii, for a smooth, underlying core mass function (see Figure \ref{fig:isocool}).

Figure \ref{fig:prim_RvP} demonstrates that the location of the primordial ``radius valley'' shifts to smaller radii farther from the star. As the disk gets colder, planet's Bondi radius increases and so the isothermal limit rises. Figure \ref{fig:isocool} illustrates this behavior where the rocky-to-enveloped transition shifts to smaller core masses at longer orbital periods. This negative slope of the valley in the radius-period space is reminiscent of that observed \citep{Fulton17,vanEylen18}. We see a larger separation between the rocky and the enveloped planetary population for dust-free gas accretion. As Figure \ref{fig:isocool} shows, this difference arises from both the generally more rapid accretion and weaker dependence on core mass for dust-free envelopes.

As we will show in the next section, gas accretion needs to be dust-free in order for the primordial radius gap (and the post-evaporation gap) to align with the observation. From numerically fitting the rocky-to-enveloped transition mass (i.e., where $M_{\rm iso}$ (equation \ref{eq:Miso}; computed numerically) crosses $M_{\rm env}$ (equation \ref{eq:df_accr})) vs.~orbital period, we find the transition mass $\propto P^{-0.31}$. Since $M_{\rm core} \propto R_{\rm core}^4$, we find the radius valley $R_{\rm valley} \propto P^{-0.08}$, in good agreement with \citet{vanEylen18} (within 1-$\sigma$ error) and \citet{Martinez19} (within 1.5-$\sigma$ error). Following the same exercise but using temperature scaling expected from active disks $T_{\rm disk} \propto a^{-3/4}$, we obtain $R_{\rm valley} \propto P^{-0.15}$ which agrees with both \citet{vanEylen18} and \citet{Martinez19} within 1.5-$\sigma$ error.

\subsection{Mass Loss and Underlying Core Mass Distribution}
\label{ssec:mass_loss}

\begin{figure*}
    \centering
    \includegraphics[width=\textwidth]{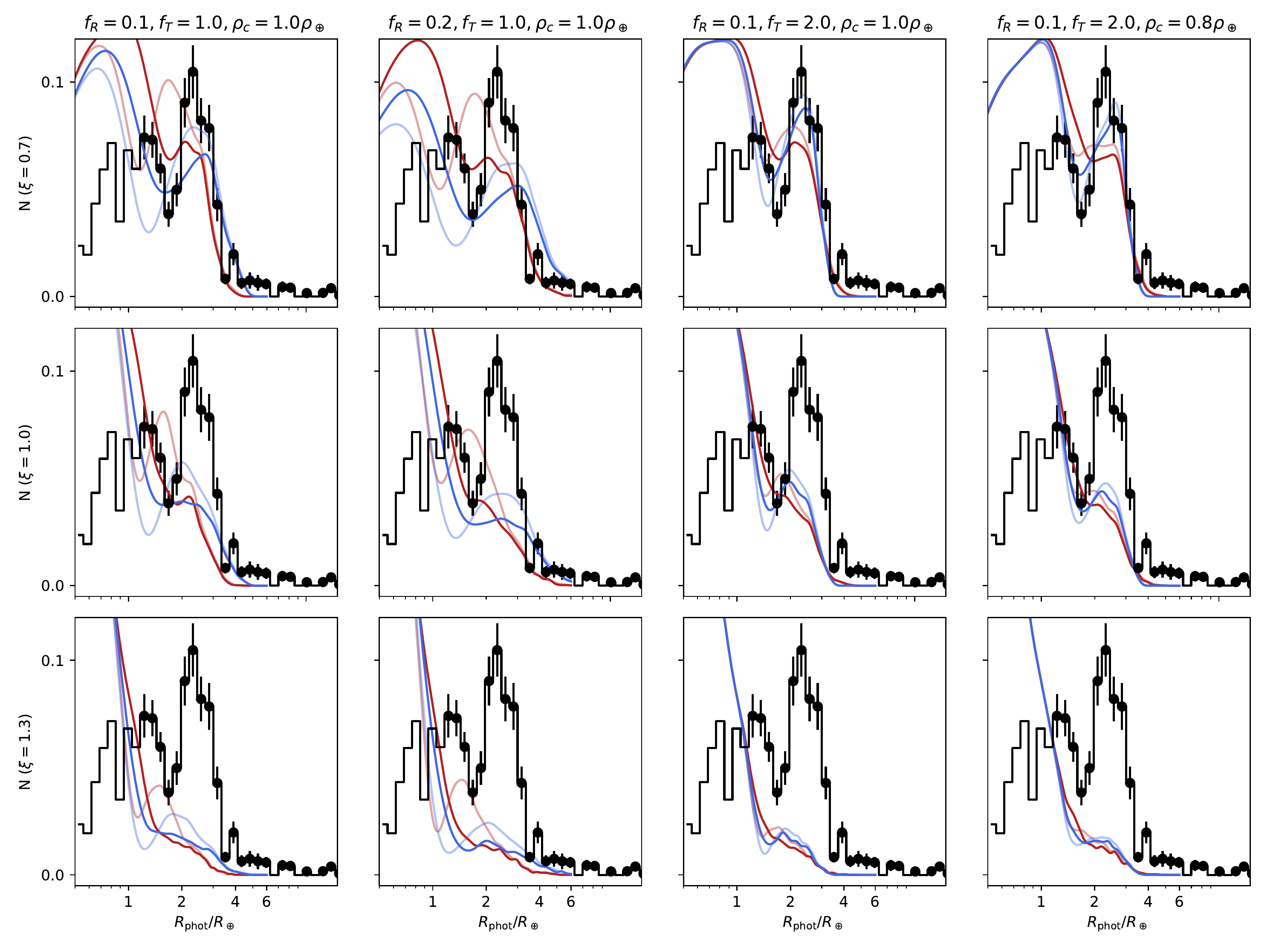}
    \caption{Distribution of planetary radii for a variety of underlying core mass distributions ($\xi$), truncation factor of the outer radius due to hydrodynamic effects ($f_R$), the disk temperature ($f_T$), and the core density ($\rho_c$). Model histograms are smoothed using Gaussian kernels with Scott's Rule for bandwidth selection (\texttt{SciPy}'s \texttt{gaussian\_kde} function). Dusty and dust-free calculations are shown in red and blue, respectively, with the primordial population drawn in lighter color while the post-evaporation populations are drawn in darker color. Data from \citet{Fulton18} are illustrated in black; data below $\sim$1$R_\oplus$ falls off their detection threshold and so the true sub-Earth population may be under-represented \citep[see, e.g.,][]{Hsu19}. In general, the location of the radius valley carved out by photoevaporation is robust to varying initial conditions while the depth and the width of the valley change considerably: hotter disks narrows the gap; larger $f_R$ broadens the overall radii distribution; fluffier cores shift the valley to larger radii; and core mass distributions that are neutral or bottom heavy are unable to reproduce the observed strong peak at $\sim$2$R_\oplus$. Among the combinations of parameters shown in this figure, dust-free envelopes with $\xi=0.7$, $f_R=0.1$, $f_T=2.0$, and $\rho_c = 0.8\rho_\oplus$ agree best with the observation.}
    \label{fig:rdistrb_param}
\end{figure*}

In the previous section, we showed how late-time gas accretion alone can produce a gap in the radius distribution and how the shape of this gap in radius-period space is in agreement with the observations.
Next, we consider how envelope mass loss further molds the final radius distribution.
Figure \ref{fig:rdistrb_param} demonstrates that the location of the radius valley carved out by photoevaporative mass loss is robustly situated at $\sim$1.6$R_\oplus$ regardless of the primordial population. As \citet{Owen17} cogently explain, gas-enveloped planets transform to bare rocky cores by photoevaporation when their envelope mass loss timescale $\lesssim$100 Myrs, the typical saturation timescale of high-energy luminosity of host stars. For our choice of parameters, this transition occurs for $M_{\rm core} \sim$ 5--6$M_\oplus$ whose core radius is $\sim$1.5--1.6$R_\oplus$ for Earth composition.

Where the initial conditions make a difference is in the depth and the width of the gap. As illustrated in Figure \ref{fig:rdistrb_param}, the narrow valley and peak in the distribution of radii are more likely to appear in dust-free envelopes (blue lines) with smaller outer radius (smaller $f_R$) that are assembled in hotter disks (higher $f_T$), and built from top-heavy core mass functions (smaller $\xi$). Cores that are slightly less dense than the Earth ($\rho_c = 0.8\rho_\oplus$) reproduce better the observed location of the radius valley as the radii of bare cores puff up slightly pushing the location of the valley to $\sim$1.7--1.8$R_\oplus$.

The narrowness of the radius peak for dust-free envelopes as opposed to dusty envelopes can be understood from the weaker dependence of $M_{\rm env}$ on $M_{\rm core}$ (see equations \ref{eq:dusty_accr} and \ref{eq:df_accr} as well as Figure \ref{fig:isocool}). For a given range of $M_{\rm core}$, the confines of possible envelope mass fractions and therefore radii are more limited. 

Smaller $f_R$ reduces the maximum $M_{\rm env}/M_{\rm core}$ and so keeps the primordial radius peak closer to the valley. Since photoevaporative mass loss effectively carves out the large radii peak and add them to the lower radii, observations are better reproduced when the initial radius valley is narrower.

In hotter disks, the isothermal maximal $M_{\rm env}/M_{\rm core}$ shrinks so that the rocky-to-enveloped transition appears at higher core masses. The result is a positive shift in the location of the primordial radius valley. The gas accretion rate for dust-free envelopes also reduces (see equation \ref{eq:df_accr}) and so the sub-Neptune peak shifts closer to $\sim$2$R_\oplus$, bringing the primordial distribution of radii in better agreement with the observation (see the faint blue line in the top third panel of Figure \ref{fig:rdistrb_param}). Since the locations of the valley are coincident with that expected from photoevaporative mass loss, we only observe slight reduction in the peak at $\sim$2$R_\oplus$ and a slight shallowing of the valley at $\sim$1.6$R_\oplus$.

We observe a loss of a peak in the radius distribution when the underlying core mass function is too bottom-heavy ($\xi = 1.3$). While we defer detailed formal fitting of models to the data for future analyses, it is already apparent that the allowed range of $\xi$ appears tightly constrained, under the ansatz that the core mass distribution follows a power-law. It may be possible to restore the radius peak even with $\xi=1.3$ with sufficiently high $f_T$ but we judge $f_T > 3$ to be unlikely as it implies the disk is hot enough to melt iron at $\sim$0.1 AU.

\begin{figure*}
    \centering
    \includegraphics[width=\textwidth]{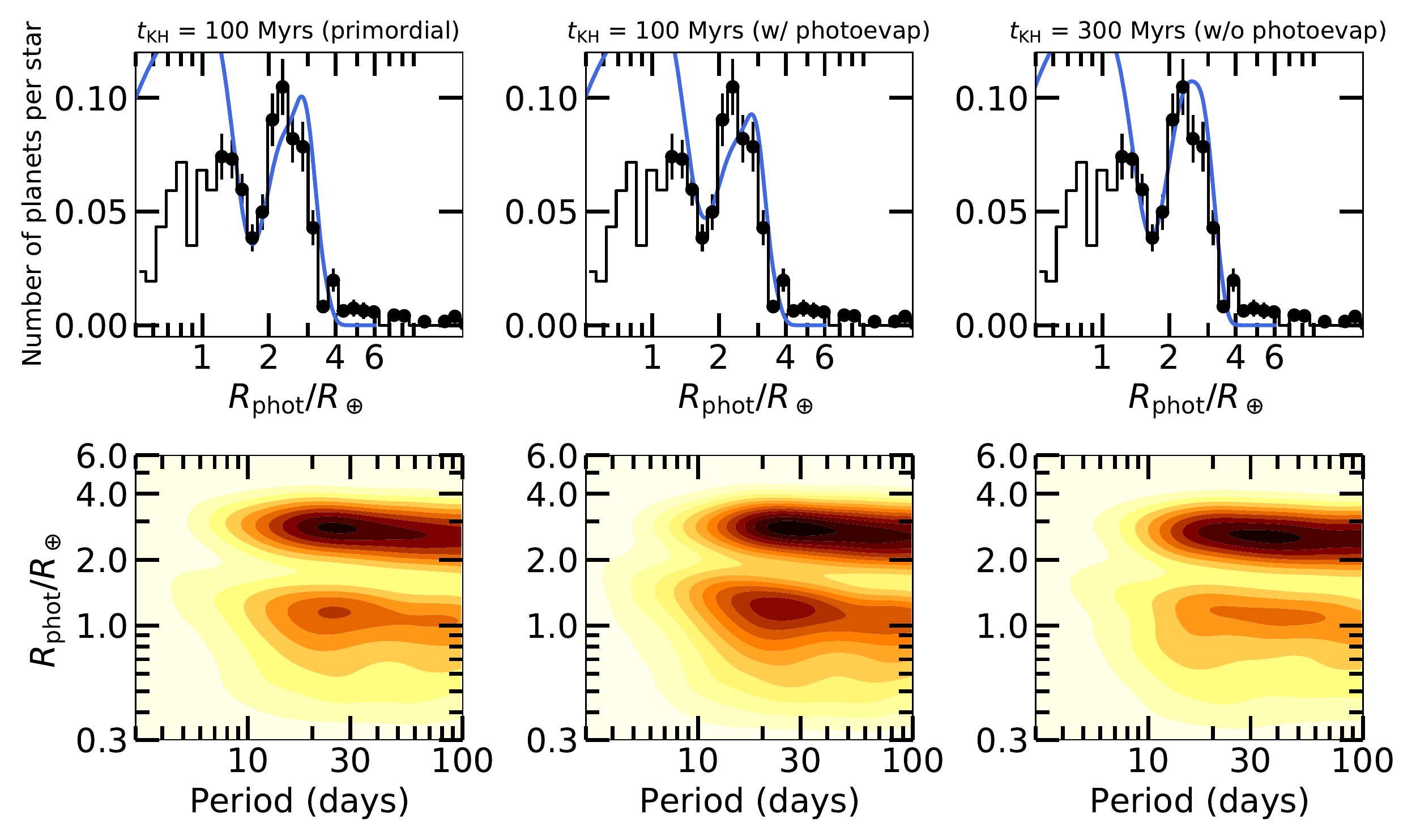}
    \caption{Radius histogram and radius-period distribution of model populations ($\xi=0.7$, $f_T=2.0$, $\rho_c=0.8\rho_\oplus$, $f_R=0.1$, $f_{\rm dep}=0.001$) that best agree with the observations. Left: primordial population of planets that assembled in gas-poor environment. Middle: population depicted in left processed by photoevaporative mass loss over 1 Gyrs. Right: planets that assembled in gas-poor environment evolved for 1 Gyrs without any mass loss. Data from \citet{Fulton18} are drawn in black. We note that data below $\sim$1$R_\oplus$ falls off their detection threshold and so the true sub-Earth population may be under-represented \citep[see, e.g.,][]{Hsu19}. Photoevaporation transforms sub-Neptunes into super-Earths effectively filling in the radius gap. The shapes of both the valley and the peak in one-dimensional radius distribution are best reproduced in the evolved model without mass loss. Quantitatively, the radius valley in the pure accretion model scales with orbital period as $R_{\rm valley} \propto P^{-0.08}$, in good agreement with \citet{vanEylen18} (within 1-$\sigma$ error) and \citet{Martinez19} (within 1.5-$\sigma$ error).}
    \label{fig:RvP_tKH}
\end{figure*}

The combinations of parameters that provides the model radius distribution agreeing best with the observation are highlighted in Figure \ref{fig:RvP_tKH}, corresponding to dust-free envelopes and $f_R, f_T, \xi, \rho_c/\rho_\oplus =$ (0.1, 2, 0.7, 0.8). Between the primordial and evaporated population, we see a closing of the gap at $\lesssim$30 days as sub-Neptunes are whittled down to bare rocky objects by photoevaporation. This transformation also fills in the valley in one-dimensional radius histogram. In fact, planet population that evolved from late time gas accretion without any subsequent mass loss (right column) appears to reproduce best the observed radius valley.

The observed radius valley closes at $\sim$10 days and widens towards $\sim$100 days \citep{Fulton18}. In photoevaporation models that assume all cores to have started with $\gtrsim 0.01$\% by mass envelope, this triangular delta is hard to reproduce if the underlying core mass function is assumed flat \citep[see, e.g.,][]{Owen13}. Accounting for core mass-dependent initial envelope mass fraction alleviates the need for peaked core mass distribution as the radius gap is already carved out and photoevaporative mass loss preferentially transforms short-period sub-Neptunes into super-Earths, closing the gap inside $\sim$30 days (see middle columen of Figure \ref{fig:RvP_tKH}).

\begin{figure*}
    \centering
    \includegraphics[width=\textwidth]{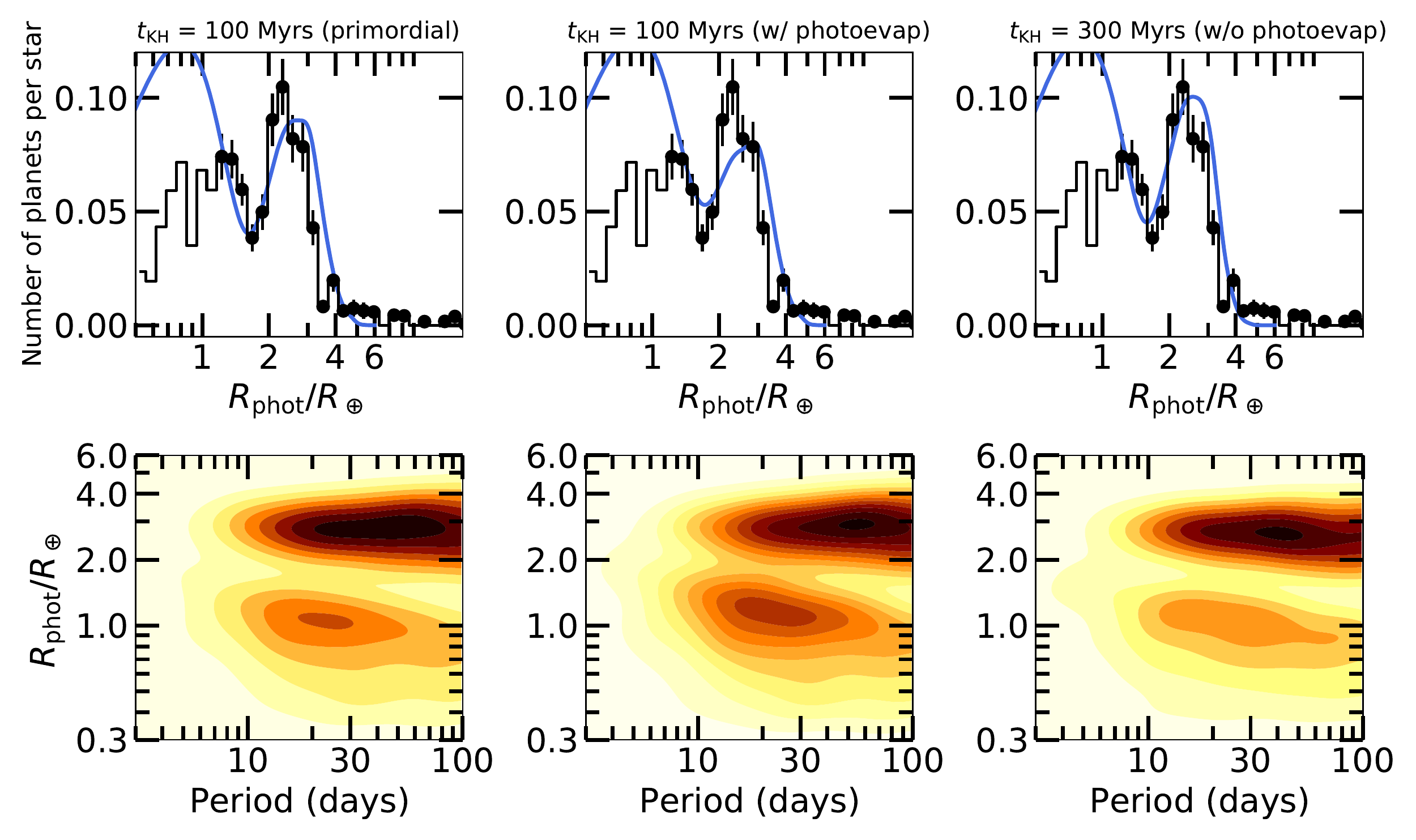}
    \caption{Same as Figure \ref{fig:RvP_tKH} except we use the disk temperature scaling that is appropriate for disks heated by viscous accretion, $T_{\rm disk} = f_T\times 1000{\rm K}(a/0.1{\rm au})^{-3/4}$. We obtain a radius valley that is steeper in the radius-period space: $R_{\rm valley} \propto P^{-0.15}$ which agrees with both \citet{vanEylen18} and \citet{Martinez19} within 1.5-$\sigma$ error.}
    \label{fig:RvP_tKH_active}
\end{figure*}

\section{Discussion and Conclusion}
\label{sec:disc_concl}

We demonstrated that the underlying core mass distribution of sub-Neptunes can be broad with substantial population of sub-Earth mass objects while still reproducing the observed gap in the radius distribution and in radius-period space. A radius gap is already in place at birth as cores lighter than $\sim$1--2$M_\oplus$ can never accrete enough gas to be observed as gas-enveloped. The maximum envelope mass given by the maximally cooled isothermal state drops exponentially with smaller core masses so that for a smooth distribution of core masses, a sharp radius dichotomy across $\sim$1--2$R_\oplus$ appears. Furthermore, this primordial radius gap shifts to smaller radii at longer orbital periods as the maximum isothermal mass rises and so the rocky-to-enveloped transition shifts to smaller cores. 

Late-time formation of sub-Neptune is often attributed to producing a positive slope of the radius-period valley, based on the calculation of \citet{Lopez18}. As these authors state, and we emphasize, their calculation is appropriate for formation in a gas-empty environment after a complete disk gas dispersal. The positive slope of the radius-period gap obtains from computing the expected core masses in a minimum mass extrasolar nebula \citep[MMEN;][]{Chiang13} which produces rising masses (and therefore radii) at larger orbital distances (using the updated MMEN by \citet{Dai20} will produce a similar result). The slope of the valley in the radius-period space may indeed turn positive around low mass stars (\citealt{Cloutier20}; but see \citealt{Wu19}).
Our premise is distinct: we consider the formation of sub-Neptunes in gas-poor but not gas-empty nebula so that gas accretion, however limited, occurs. It is formation that is late-time in terms of the evolution of disk gas but not so late that there is no gas left (e.g., inner holes of transitional disks).

Our model of late-time gas accretion best reproduces the observed location, width, and depth of the radius gap when the sub-Neptune cores follow mass functions shallower than $dN/dM_{\rm core} \propto M_{\rm core}^{-1}$ and accrete dust-free gas in hot disks.\footnote{We find a potentially good agreement using a bottom-heavy core mass function $dN/dM_{\rm core} \propto M_{\rm core}^{-1.1}$ for puffy cores but only in one-dimensional radius histogram. The triangular shape of the radius-period valley is challenging to reproduce with bottom-heavy core mass functions.} The rate of accretion in dusty environment is too sensitive to core mass so that the final distribution of envelope mass fractions and therefore radii is too broad compared to that observed. The coagulation and the rain-out of dust grains may be an efficient process in sub-Neptune envelopes \citep{Ormel14}.

\subsection{Dependence on Disk Temperature and Stellar Mass}
\label{ssec:Tdisk}

The required disk temperatures may be uncomfortably high. We note that for accretion disks, the mid-plane temperature at $\sim$0.1 AU can be as high as $\sim$2000K \citep[see][their Figure 3]{Dalessio98}, consistent with our $f_T=2$. \citet{Dalessio98} accounted for both accretional and irradiation heating where the former tends to dominate in the inner disk. The expected midplane temperature from accretion depends on the accretion rate to the power of 1/4. \citet{Dalessio98} adopted the typical T Tauri accretion rate $\dot{M} = 10^{-8} M_\odot\,{\rm yr}^{-1}$. With $f_\rho = 0.001$ with respect to MMEN that we adopted here, this accretion rate can be attained with Shakura-Sunyaev viscous parameter $\alpha \sim 8\times 10^{-3}$ which is larger than what is usually inferred from studies of turbulence in protoplanetary disks such as HL tau \citep{Pinte16}, but it is below the largest detected level of turbulence in e.g., DM Tau \citep{Flaherty20}. For optically thick disks, the midplane temperature further enhances by a factor $\tau^{1/4}$ where $\tau$ is the vertical optical depth. \citet{Dalessio98} compute $\tau \sim 400$ at 0.1 AU (see their Figure 2) which can be accommodated with $f_\rho = 0.001$ assuming well-mixed dust-gas mixture within the disk \citep[see][their Figure 1]{Lee18}. (This enhancement of midplane temperature by $\tau^{1/4}$ applies to passive disks as well.) Gas accretion onto planets can still proceed in dust-free manner as long as the dust grains coagulate and settle within the planetary atmosphere \citep[e.g.,][]{Ormel14}. Figure \ref{fig:RvP_tKH_active} demonstrates that qualitative result---that late-time gas accretion alone can reproduce the radius valley at the observed location---holds when we explicitly adopt the disk temperature scaling that is appropriate for disk heated by viscous accretion (i.e., active disks).

Matching the location of the primordial radius valley with that observed requires that the sub-Neptune cores are slightly less dense, e.g., $\sim$80\% of the Earth composition, in rough agreement with what is reported for short-period super-Earths by \citet{Dorn19} and more generally by \citet{Rogers20}. We note that some of the model parameters that produce a broad peak may be narrowed by taking into account core-envelope interaction, in particular, the dissolution of gas into the magma core \citep{Kite19}. Assessing the effect of core-envelope mixing at formation is a subject of our ongoing work.

We note that the primordial radius valley is expected to shift towards larger sizes around higher mass stars, assuming their disks are hotter. For disks heated by stellar irradiation, $T_{\rm disk} \propto T_{\rm eff} (R_\star/a)^{1/2}$ where $T_{\rm eff} \propto M_\star^{1/7}$ is the effective temperature of the star and $R_\star \propto M_\star^{1/2}$ is the radius of the star, all evaluated for fully convective, pre-main sequence stars \citep[e.g.,][]{Kippenhahn}.\footnote{Mass-radius relationship for fully convective pre-main sequence star can be obtained by evaluating time evolution of radius with luminosity given by convection: $R_\star \propto t^{-1/3}M_\star^{1/2}$.} For these passive disks, $T_{\rm disk} \propto M_\star^{11/28}$. From numerical fitting, we find the rocky-to-gas-enveloped transition core mass $M_{\rm c,trans} \propto T_{\rm disk}^{1.09}$ and $R_c \propto M_c^{1/4}$ so the radius valley $R_{\rm val} \propto M_\star^{0.11}$. For disks heated by accretion, $T_{\rm disk} \propto (M_\star \dot{M}/R_\star^3)^{1/4}(a/R_\star)^{-3/4}$ \citep{Clarke}. Taking $\dot{M} \propto M_\star^{1.95}$ \citep{Calvet04}, we find $T_{\rm disk} \propto M_\star^{0.7}$. Again, from numerical fitting but for active disks, $M_{\rm c,trans} \propto T_{\rm disk}^{2.10}$ and so $R_{\rm val} \propto M_\star^{0.37}$. Both estimates are within the 1-$\sigma$ error bar estimate from Gaia-Kepler catalog by \citet{Berger20}.

We conclude this section by noting that computing disk temperature depends on uncertain factors including dust grain size distribution that can vary both radially and vertically. The disk temperatures we adopted in this paper are in the realm of possibility but more accurate comparison with observations will require better understanding of the thermal structure of the protoplanetary disks and their dependence on the host stellar mass.

\subsection{Primordial vs.~Mass Loss}

Late-time gas accretion alone can carve out a valley in radius-period distribution of exoplanets. 
In fact, in many of the cases we explore, the agreement with the data becomes worse after taking into account photoevaporative mass loss, as the radius gap fills up. Intriguingly, \citet{David20} report the radius gap appears more filled in around stars older than $\sim$2 Gyrs, potentially consistent with our results as long as the dominant mass loss process operates over 1--2 Gyrs which can be either photoevaporation \citep[e.g.,][]{King20} or core-powered mass loss \citep[e.g.,][]{Ginzburg18} (or both). We note however that \citet{Berger20} report no discernible change in the depth of the gap across stellar ages of $\sim$1 Gyr. What the two studies have in common is that the relative number fraction of super-Earth rises around older stars, suggesting long-term mass loss processes are likely in effect.


There remain uncertainties in the exact magnitude of the mass loss for both photoevaporation and core-powered mass loss models. In the picture of photoevaporation, the unknown strength of planetary magnetic fields can shield against high-energy stellar photons \citep{Owen19}. Furthermore, there is an order of magnitude variation in the magnitude and time evolution of stellar EUV and X-ray luminosity \citep{Tu15}. In the picture of core-powered envelope mass loss, the amount of gas mass that can be lost via wind depends on the structure of the outer envelope subject to uncertain opacity sources. Even if the cores hold enough thermal energy to unbind the entire envelope, the timescale of heat transfer depends on the unknown viscosity and Prandtl number of the magma/rocky core \citep[e.g.,][]{Stamenkovic12,Garaud18,Fuentes20}. Further advances in both theory and observations should iron out these uncertainties. 

\acknowledgements
The anonymous referee provided a thoughtful report that helped improve the manuscript. We thank Ruth Murray-Clay for her encouragement to pursue this work, as well as BJ Fulton, Erik Petigura, and Tony Piro for motivating us to look more closely at the radius distribution expected from late-time gas accretion. We also thank Eugene Chiang, Ryan Cloutier, James Owen, and Yanqin Wu for helpful discussions. EJL is supported by the Natural Sciences and Engineering Research Council of Canada (NSERC), RGPIN-2020-07045.

\bibliography{isocores}{}

\begin{thebibliography}{}
\expandafter\ifx\csname natexlab\endcsname\relax\def\natexlab#1{#1}\fi
\providecommand{\url}[1]{\href{#1}{#1}}
\providecommand{\dodoi}[1]{doi:~\href{http://doi.org/#1}{\nolinkurl{#1}}}
\providecommand{\doeprint}[1]{\href{http://ascl.net/#1}{\nolinkurl{http://ascl.net/#1}}}
\providecommand{\doarXiv}[1]{\href{https://arxiv.org/abs/#1}{\nolinkurl{https://arxiv.org/abs/#1}}}

\bibitem[{{Ali-Dib} {et~al.}(2020){Ali-Dib}, {Cumming}, \& {Lin}}]{Alidib20}
{Ali-Dib}, M., {Cumming}, A., \& {Lin}, D. N.~C. 2020, \mnras, 494, 2440,
  \dodoi{10.1093/mnras/staa914}

\bibitem[{{Berger} {et~al.}(2020){Berger}, {Huber}, {Gaidos}, {van Saders}, \&
  {Weiss}}]{Berger20}
{Berger}, T.~A., {Huber}, D., {Gaidos}, E., {van Saders}, J.~L., \& {Weiss},
  L.~M. 2020, arXiv e-prints, arXiv:2005.14671.
\newblock \doarXiv{2005.14671}

\bibitem[{{Burke} {et~al.}(2015){Burke}, {Christiansen}, {Mullally}, {Seader},
  {Huber}, {Rowe}, {Coughlin}, {Thompson}, {Catanzarite}, {Clarke}, {Morton},
  {Caldwell}, {Bryson}, {Haas}, {Batalha}, {Jenkins}, {Tenenbaum}, {Twicken},
  {Li}, {Quintana}, {Barclay}, {Henze}, {Borucki}, {Howell}, \&
  {Still}}]{Burke15}
{Burke}, C.~J., {Christiansen}, J.~L., {Mullally}, F., {et~al.} 2015, \apj,
  809, 8, \dodoi{10.1088/0004-637X/809/1/8}

\bibitem[{{Calvet} {et~al.}(2004){Calvet}, {Muzerolle}, {Brice{\~n}o},
  {Hern{\'a}ndez}, {Hartmann}, {Saucedo}, \& {Gordon}}]{Calvet04}
{Calvet}, N., {Muzerolle}, J., {Brice{\~n}o}, C., {et~al.} 2004, \aj, 128,
  1294, \dodoi{10.1086/422733}

\bibitem[{{Chiang} \& {Laughlin}(2013)}]{Chiang13}
{Chiang}, E., \& {Laughlin}, G. 2013, \mnras, 431, 3444,
  \dodoi{10.1093/mnras/stt424}

\bibitem[{{Choksi} \& {Chiang}(2020)}]{Choksi20}
{Choksi}, N., \& {Chiang}, E. 2020, \mnras, 495, 4192,
  \dodoi{10.1093/mnras/staa1421}

\bibitem[{{Clarke} \& {Carswell}(2007)}]{Clarke}
{Clarke}, C., \& {Carswell}, B. 2007, {Principles of Astrophysical Fluid
  Dynamics}

\bibitem[{{Cloutier} \& {Menou}(2020)}]{Cloutier20}
{Cloutier}, R., \& {Menou}, K. 2020, \aj, 159, 211,
  \dodoi{10.3847/1538-3881/ab8237}

\bibitem[{{Dai} {et~al.}(2020){Dai}, {Winn}, {Schlaufman}, {Wang}, {Weiss},
  {Petigura}, {Howard}, \& {Fang}}]{Dai20}
{Dai}, F., {Winn}, J.~N., {Schlaufman}, K., {et~al.} 2020, \aj, 159, 247,
  \dodoi{10.3847/1538-3881/ab88b8}

\bibitem[{{D'Alessio} {et~al.}(1998){D'Alessio}, {Cant{\"o}}, {Calvet}, \&
  {Lizano}}]{Dalessio98}
{D'Alessio}, P., {Cant{\"o}}, J., {Calvet}, N., \& {Lizano}, S. 1998, \apj,
  500, 411, \dodoi{10.1086/305702}

\bibitem[{{David} {et~al.}(2020){David}, {Contardo}, {Sandoval}, {Angus},
  {Yuxi}, {Lu}, {Bedell}, {Curtis}, {Foreman-Mackey}, {Fulton}, {Grunblatt}, \&
  {Petigura}}]{David20}
{David}, T.~J., {Contardo}, G., {Sandoval}, A., {et~al.} 2020, arXiv e-prints,
  arXiv:2011.09894.
\newblock \doarXiv{2011.09894}

\bibitem[{{Dawson} {et~al.}(2015){Dawson}, {Chiang}, \& {Lee}}]{Dawson15}
{Dawson}, R.~I., {Chiang}, E., \& {Lee}, E.~J. 2015, \mnras, 453, 1471,
  \dodoi{10.1093/mnras/stv1639}

\bibitem[{{Dorn} {et~al.}(2019){Dorn}, {Harrison}, {Bonsor}, \&
  {Hands}}]{Dorn19}
{Dorn}, C., {Harrison}, J.~H.~D., {Bonsor}, A., \& {Hands}, T.~O. 2019, \mnras,
  484, 712, \dodoi{10.1093/mnras/sty3435}

\bibitem[{{Flaherty} {et~al.}(2020){Flaherty}, {Hughes}, {Simon}, {Qi}, {Bai},
  {Bulatek}, {Andrews}, {Wilner}, \& {K{\'o}sp{\'a}l}}]{Flaherty20}
{Flaherty}, K., {Hughes}, A.~M., {Simon}, J.~B., {et~al.} 2020, \apj, 895, 109,
  \dodoi{10.3847/1538-4357/ab8cc5}

\bibitem[{{Freedman} {et~al.}(2008){Freedman}, {Marley}, \&
  {Lodders}}]{Freedman08}
{Freedman}, R.~S., {Marley}, M.~S., \& {Lodders}, K. 2008, \apjs, 174, 504,
  \dodoi{10.1086/521793}

\bibitem[{{Fressin} {et~al.}(2013){Fressin}, {Torres}, {Charbonneau}, {Bryson},
  {Christiansen}, {Dressing}, {Jenkins}, {Walkowicz}, \& {Batalha}}]{Fressin13}
{Fressin}, F., {Torres}, G., {Charbonneau}, D., {et~al.} 2013, \apj, 766, 81,
  \dodoi{10.1088/0004-637X/766/2/81}

\bibitem[{{Fuentes} \& {Cumming}(2020)}]{Fuentes20}
{Fuentes}, J.~R., \& {Cumming}, A. 2020, arXiv e-prints, arXiv:2007.04265.
\newblock \doarXiv{2007.04265}

\bibitem[{{Fulton} \& {Petigura}(2018)}]{Fulton18}
{Fulton}, B.~J., \& {Petigura}, E.~A. 2018, \aj, 156, 264,
  \dodoi{10.3847/1538-3881/aae828}

\bibitem[{{Fulton} {et~al.}(2017){Fulton}, {Petigura}, {Howard}, {Isaacson},
  {Marcy}, {Cargile}, {Hebb}, {Weiss}, {Johnson}, {Morton}, {Sinukoff},
  {Crossfield}, \& {Hirsch}}]{Fulton17}
{Fulton}, B.~J., {Petigura}, E.~A., {Howard}, A.~W., {et~al.} 2017, \aj, 154,
  109, \dodoi{10.3847/1538-3881/aa80eb}

\bibitem[{{Fung} {et~al.}(2019){Fung}, {Zhu}, \& {Chiang}}]{Fung19}
{Fung}, J., {Zhu}, Z., \& {Chiang}, E. 2019, \apj, 887, 152,
  \dodoi{10.3847/1538-4357/ab53da}

\bibitem[{{Garaud}(2018)}]{Garaud18}
{Garaud}, P. 2018, Annual Review of Fluid Mechanics, 50, 275,
  \dodoi{10.1146/annurev-fluid-122316-045234}

\bibitem[{{Ginzburg} {et~al.}(2016){Ginzburg}, {Schlichting}, \&
  {Sari}}]{Ginzburg16}
{Ginzburg}, S., {Schlichting}, H.~E., \& {Sari}, R. 2016, \apj, 825, 29,
  \dodoi{10.3847/0004-637X/825/1/29}

\bibitem[{{Ginzburg} {et~al.}(2018){Ginzburg}, {Schlichting}, \&
  {Sari}}]{Ginzburg18}
---. 2018, \mnras, 476, 759, \dodoi{10.1093/mnras/sty290}

\bibitem[{{Gupta} \& {Schlichting}(2020)}]{Gupta20}
{Gupta}, A., \& {Schlichting}, H.~E. 2020, \mnras, 493, 792,
  \dodoi{10.1093/mnras/staa315}

\bibitem[{{Hadden} \& {Lithwick}(2014)}]{Hadden14}
{Hadden}, S., \& {Lithwick}, Y. 2014, \apj, 787, 80,
  \dodoi{10.1088/0004-637X/787/1/80}

\bibitem[{{Hsu} {et~al.}(2019){Hsu}, {Ford}, {Ragozzine}, \& {Ashby}}]{Hsu19}
{Hsu}, D.~C., {Ford}, E.~B., {Ragozzine}, D., \& {Ashby}, K. 2019, \aj, 158,
  109, \dodoi{10.3847/1538-3881/ab31ab}

\bibitem[{{Ikoma} \& {Hori}(2012)}]{Ikoma12}
{Ikoma}, M., \& {Hori}, Y. 2012, \apj, 753, 66,
  \dodoi{10.1088/0004-637X/753/1/66}

\bibitem[{{Jackson} {et~al.}(2012){Jackson}, {Davis}, \&
  {Wheatley}}]{Jackson12}
{Jackson}, A.~P., {Davis}, T.~A., \& {Wheatley}, P.~J. 2012, \mnras, 422, 2024,
  \dodoi{10.1111/j.1365-2966.2012.20657.x}

\bibitem[{{King} \& {Wheatley}(2020)}]{King20}
{King}, G.~W., \& {Wheatley}, P.~J. 2020, arXiv e-prints, arXiv:2007.13731.
\newblock \doarXiv{2007.13731}

\bibitem[{{Kippenhahn} {et~al.}(2012){Kippenhahn}, {Weigert}, \&
  {Weiss}}]{Kippenhahn}
{Kippenhahn}, R., {Weigert}, A., \& {Weiss}, A. 2012, {Stellar Structure and
  Evolution}, \dodoi{10.1007/978-3-642-30304-3}

\bibitem[{{Kite} {et~al.}(2019){Kite}, {Fegley}, {Schaefer}, \&
  {Ford}}]{Kite19}
{Kite}, E.~S., {Fegley}, Bruce, J., {Schaefer}, L., \& {Ford}, E.~B. 2019,
  \apjl, 887, L33, \dodoi{10.3847/2041-8213/ab59d9}

\bibitem[{{Lambrechts} \& {Lega}(2017)}]{Lambrechts17}
{Lambrechts}, M., \& {Lega}, E. 2017, \aap, 606, A146,
  \dodoi{10.1051/0004-6361/201731014}

\bibitem[{{Lee} \& {Chiang}(2015)}]{Lee15}
{Lee}, E.~J., \& {Chiang}, E. 2015, \apj, 811, 41,
  \dodoi{10.1088/0004-637X/811/1/41}

\bibitem[{{Lee} \& {Chiang}(2016)}]{Lee16}
---. 2016, \apj, 817, 90, \dodoi{10.3847/0004-637X/817/2/90}

\bibitem[{{Lee} {et~al.}(2018){Lee}, {Chiang}, \& {Ferguson}}]{Lee18}
{Lee}, E.~J., {Chiang}, E., \& {Ferguson}, J.~W. 2018, \mnras, 476, 2199,
  \dodoi{10.1093/mnras/sty389}

\bibitem[{{Lee} {et~al.}(2014){Lee}, {Chiang}, \& {Ormel}}]{Lee14}
{Lee}, E.~J., {Chiang}, E., \& {Ormel}, C.~W. 2014, \apj, 797, 95,
  \dodoi{10.1088/0004-637X/797/2/95}

\bibitem[{{Lopez} \& {Fortney}(2013)}]{Lopez13}
{Lopez}, E.~D., \& {Fortney}, J.~J. 2013, \apj, 776, 2,
  \dodoi{10.1088/0004-637X/776/1/2}

\bibitem[{{Lopez} \& {Fortney}(2014)}]{Lopez14}
---. 2014, \apj, 792, 1, \dodoi{10.1088/0004-637X/792/1/1}

\bibitem[{{Lopez} \& {Rice}(2018)}]{Lopez18}
{Lopez}, E.~D., \& {Rice}, K. 2018, \mnras, 479, 5303,
  \dodoi{10.1093/mnras/sty1707}

\bibitem[{{Malhotra}(2015)}]{Malhotra15}
{Malhotra}, R. 2015, \apj, 808, 71, \dodoi{10.1088/0004-637X/808/1/71}

\bibitem[{{Marcy} {et~al.}(2014){Marcy}, {Isaacson}, {Howard}, {Rowe},
  {Jenkins}, {Bryson}, {Latham}, {Howell}, {Gautier}, {Batalha}, {Rogers},
  {Ciardi}, {Fischer}, {Gilliland}, {Kjeldsen}, {Christensen-Dalsgaard},
  {Huber}, {Chaplin}, {Basu}, {Buchhave}, {Quinn}, {Borucki}, {Koch}, {Hunter},
  {Caldwell}, {Van Cleve}, {Kolbl}, {Weiss}, {Petigura}, {Seager}, {Morton},
  {Johnson}, {Ballard}, {Burke}, {Cochran}, {Endl}, {MacQueen}, {Everett},
  {Lissauer}, {Ford}, {Torres}, {Fressin}, {Brown}, {Steffen}, {Charbonneau},
  {Basri}, {Sasselov}, {Winn}, {Sanchis-Ojeda}, {Christiansen}, {Adams},
  {Henze}, {Dupree}, {Fabrycky}, {Fortney}, {Tarter}, {Holman}, {Tenenbaum},
  {Shporer}, {Lucas}, {Welsh}, {Orosz}, {Bedding}, {Campante}, {Davies},
  {Elsworth}, {Handberg}, {Hekker}, {Karoff}, {Kawaler}, {Lund}, {Lundkvist},
  {Metcalfe}, {Miglio}, {Silva Aguirre}, {Stello}, {White}, {Boss}, {Devore},
  {Gould}, {Prsa}, {Agol}, {Barclay}, {Coughlin}, {Brugamyer}, {Mullally},
  {Quintana}, {Still}, {Thompson}, {Morrison}, {Twicken}, {D{\'e}sert},
  {Carter}, {Crepp}, {H{\'e}brard}, {Santerne}, {Moutou}, {Sobeck}, {Hudgins},
  {Haas}, {Robertson}, {Lillo-Box}, \& {Barrado}}]{Marcy14}
{Marcy}, G.~W., {Isaacson}, H., {Howard}, A.~W., {et~al.} 2014, \apjs, 210, 20,
  \dodoi{10.1088/0067-0049/210/2/20}

\bibitem[{{Martinez} {et~al.}(2019){Martinez}, {Cunha}, {Ghezzi}, \&
  {Smith}}]{Martinez19}
{Martinez}, C.~F., {Cunha}, K., {Ghezzi}, L., \& {Smith}, V.~V. 2019, \apj,
  875, 29, \dodoi{10.3847/1538-4357/ab0d93}

\bibitem[{{Ormel}(2014)}]{Ormel14}
{Ormel}, C.~W. 2014, \apjl, 789, L18, \dodoi{10.1088/2041-8205/789/1/L18}

\bibitem[{{Owen} \& {Adams}(2019)}]{Owen19}
{Owen}, J.~E., \& {Adams}, F.~C. 2019, \mnras, 490, 15,
  \dodoi{10.1093/mnras/stz2601}

\bibitem[{{Owen} \& {Wu}(2013)}]{Owen13}
{Owen}, J.~E., \& {Wu}, Y. 2013, \apj, 775, 105,
  \dodoi{10.1088/0004-637X/775/2/105}

\bibitem[{{Owen} \& {Wu}(2016)}]{Owen16}
---. 2016, \apj, 817, 107, \dodoi{10.3847/0004-637X/817/2/107}

\bibitem[{{Owen} \& {Wu}(2017)}]{Owen17}
---. 2017, \apj, 847, 29, \dodoi{10.3847/1538-4357/aa890a}

\bibitem[{{Petigura} {et~al.}(2013){Petigura}, {Howard}, \&
  {Marcy}}]{Petigura13}
{Petigura}, E.~A., {Howard}, A.~W., \& {Marcy}, G.~W. 2013, Proceedings of the
  National Academy of Science, 110, 19273, \dodoi{10.1073/pnas.1319909110}

\bibitem[{{Petigura} {et~al.}(2018){Petigura}, {Marcy}, {Winn}, {Weiss},
  {Fulton}, {Howard}, {Sinukoff}, {Isaacson}, {Morton}, \&
  {Johnson}}]{Petigura18}
{Petigura}, E.~A., {Marcy}, G.~W., {Winn}, J.~N., {et~al.} 2018, \aj, 155, 89,
  \dodoi{10.3847/1538-3881/aaa54c}

\bibitem[{{Pinte} {et~al.}(2016){Pinte}, {Dent}, {M{\'e}nard}, {Hales}, {Hill},
  {Cortes}, \& {de Gregorio-Monsalvo}}]{Pinte16}
{Pinte}, C., {Dent}, W.~R.~F., {M{\'e}nard}, F., {et~al.} 2016, \apj, 816, 25,
  \dodoi{10.3847/0004-637X/816/1/25}

\bibitem[{{Pollack} {et~al.}(1996){Pollack}, {Hubickyj}, {Bodenheimer},
  {Lissauer}, {Podolak}, \& {Greenzweig}}]{Pollack96}
{Pollack}, J.~B., {Hubickyj}, O., {Bodenheimer}, P., {et~al.} 1996, \icarus,
  124, 62, \dodoi{10.1006/icar.1996.0190}

\bibitem[{{Ribas} {et~al.}(2005){Ribas}, {Guinan}, {G{\"u}del}, \&
  {Audard}}]{Ribas05}
{Ribas}, I., {Guinan}, E.~F., {G{\"u}del}, M., \& {Audard}, M. 2005, \apj, 622,
  680, \dodoi{10.1086/427977}

\bibitem[{{Rogers} \& {Owen}(2020)}]{Rogers20}
{Rogers}, J.~G., \& {Owen}, J.~E. 2020, arXiv e-prints, arXiv:2007.11006.
\newblock \doarXiv{2007.11006}

\bibitem[{{Rogers} \& {Seager}(2010)}]{Rogers10}
{Rogers}, L.~A., \& {Seager}, S. 2010, \apj, 712, 974,
  \dodoi{10.1088/0004-637X/712/2/974}

\bibitem[{{Stamenkovi{\'c}} {et~al.}(2012){Stamenkovi{\'c}}, {Noack}, {Breuer},
  \& {Spohn}}]{Stamenkovic12}
{Stamenkovi{\'c}}, V., {Noack}, L., {Breuer}, D., \& {Spohn}, T. 2012, \apj,
  748, 41, \dodoi{10.1088/0004-637X/748/1/41}

\bibitem[{{Tu} {et~al.}(2015){Tu}, {Johnstone}, {G{\"u}del}, \&
  {Lammer}}]{Tu15}
{Tu}, L., {Johnstone}, C.~P., {G{\"u}del}, M., \& {Lammer}, H. 2015, \aap, 577,
  L3, \dodoi{10.1051/0004-6361/201526146}

\bibitem[{{Valencia} {et~al.}(2006){Valencia}, {O'Connell}, \&
  {Sasselov}}]{Valencia06}
{Valencia}, D., {O'Connell}, R.~J., \& {Sasselov}, D. 2006, \icarus, 181, 545,
  \dodoi{10.1016/j.icarus.2005.11.021}

\bibitem[{{Van Eylen} {et~al.}(2018){Van Eylen}, {Agentoft}, {Lundkvist},
  {Kjeldsen}, {Owen}, {Fulton}, {Petigura}, \& {Snellen}}]{vanEylen18}
{Van Eylen}, V., {Agentoft}, C., {Lundkvist}, M.~S., {et~al.} 2018, \mnras,
  479, 4786, \dodoi{10.1093/mnras/sty1783}

\bibitem[{{Weiss} \& {Marcy}(2014)}]{Weiss14}
{Weiss}, L.~M., \& {Marcy}, G.~W. 2014, \apjl, 783, L6,
  \dodoi{10.1088/2041-8205/783/1/L6}

\bibitem[{{Wu}(2019)}]{Wu19}
{Wu}, Y. 2019, \apj, 874, 91, \dodoi{10.3847/1538-4357/ab06f8}

\bibitem[{{Wu} \& {Lithwick}(2013)}]{Wu-ttv13}
{Wu}, Y., \& {Lithwick}, Y. 2013, \apj, 772, 74,
  \dodoi{10.1088/0004-637X/772/1/74}

\end{thebibliography}
\bibliographystyle{aasjournal}

\end{document}